\documentclass{article}
\usepackage{graphicx}
\usepackage{amsmath}

\input{tcilatex}

\begin{document}

\title{Gordian unknots}
\author{P. Pieranski$^{1}$, S. Przybyl$^{1}$ and A. Stasiak$^{2}$ \\
$^{1}$Poznan University of Technology\\
e-mail: Piotr.Pieranski@put.poznan.pl\\
Nieszawska 13A, 60 965 Poznan, Poland\\
$^{2}$University of Lausanne, Switzerland}
\maketitle

\begin{abstract}
Numerical simulations indicate that there exist conformations of the unknot,
tied on a finite piece of rope, entangled in such a manner, that they cannot
be disentangled to the torus conformation without cutting the rope. The
simplest example of such a gordian unknot is presented.
\end{abstract}

Knots are closed, self-avoiding curves in the 3-dimensional space. The shape
and size of a knot, i.e. its conformation, can be changed in a very broad
range without changing the knot type. The necessary condition to keep the
knot type intact is that during all transformations applied to the knot the
curve must remain self-avoiding. From the topological point of view, all
conformations of a knot are equivalent but if the knot is considered as a
physical object, it may be not so. Let us give a simple example. Take a
concrete, knotted space curve $K$. Imagine, that $K$ is inflated into a tube
of diameter $D$. If $K$ is scaled down without scaling down $D$, then there
is obviously a minimum size below which one cannot go without changing the
shape of $K$. Diminishing, in a thought or computer experiment, the size of
a knot one arrives to the limit below which in some places of the knot the
impenetrability of the tube on which it has been tied would be violated.

Consider a knot tied on a piece of a rope. If the knot is tied in a loose
manner, one can easily change its shape. However, the range of
transformations available in such a process is much more narrow than in the
case of knots tied on an infinitely thin rope. Limitations imposed on the
transformations used to change the knot shape by the fixed thickness and
length of the rope may make some conformations of the knot inaccessible from
each other. The limitations can be in an elegant manner represented by the
single condition that the global curvature of the knot cannot be larger than 
$2/D$ \cite{Gonzalez99}. That it is the case we shall try to demonstrate in
the most simple case of the unknot. The knot is a particular one since we
know for it the shape of the ideal, least rope consuming conformation \cite
{Katritch96}. The simplest shape of the unknot is obviously circular. If the
knot is tied on the rope of diameter $D$ the shortest piece of rope one must
use to form it has the length $L_{min}=\pi D$. If one starts from the
circular conformation of the unknot tied on a longer piece of rope, the
length of the rope can be subsequently reduced without changing the circular
shape until the $L_{min}$ value is reached.

Consider now a different, entangled conformation of the unknot tied on a
piece of rope having the length $L>L_{min}$. Can it be disentangled to the
canonical circular shape? Are there such conformations of the unknot, which
cannot be disentangled to a circle without elongating the rope? For obvious
reasons we propose to call such conformations gordian. In what follows we
shall report results of numerical experiments suggesting existence of the
gordian conformations of the unknot.

\FRAME{ftbpFU}{3.7922in}{2.5486in}{0pt}{\Qcb{SONO\ disentagles an unknot
entagled in a simple manner. How the length of the rope changes in this
process is shown in Fig.2 (lower curve).}}{\Qlb{Fig1}}{fig1.jpg}{\special%
{language "Scientific Word";type "GRAPHIC";maintain-aspect-ratio
TRUE;display "USEDEF";valid_file "F";width 3.7922in;height 2.5486in;depth
0pt;original-width 8.8539in;original-height 5.9378in;cropleft "0";croptop
"1";cropright "1";cropbottom "0";filename '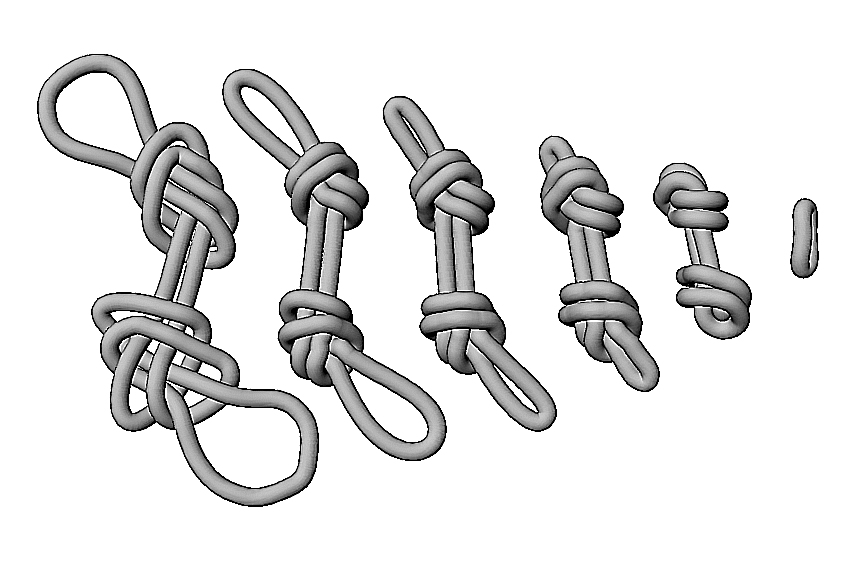';file-properties
"XNPEU";}}

Imagine that the entangled conformation of the unknot is tied on piece the
ideal rope of diameter $D$ and length $L>Lmin$. The ideal rope is perfectly
flexible but at the same time perfectly hard. Its perpendicular
cross-sections remain always circular. The diameters of all the
cross-sections are equal $D$. None of the circular cross-sections overlap.
The surface of the rope is perfectly slippery. In such conditions one may
try to force the knot to disentangle itself just by shortening the rope
length. Such a process, in which the knot is tightened, can be easily
simulated with a computer. The details of SONO (Shrink-On-No-Overlaps), the
simulation algorithm we developed, are described elsewhere\cite{MyChapter}.
As shown in \cite{MyChapter}, SONO disentangles some simple conformations of
the unknot. See Fig.\ref{Fig1}. It manages to cope also with the more
complex conformation proposed by Freedman \cite{Freedman94} disentangled
previously by the Kusner and Sullivan algorithm minimizing the M\"{o}bius
energy \cite{Kusner98}.

The steps of the construction of the Freedman conformation, are as follows
\cite{Hass}:

1. Take a circular unknot and splash it into a flat double rope band.

2. Tie overhand knots on both ends of the band and tighten them. (From the
point of view of the knot theory, the overhand knots are open trefoil knots.)

3. Open and slip the end loops over the bodies of the overhand knots, so
that they meet in the central part of the band.

4. Move the rope through both overhand knots so that the loops become
smaller.

In what follows we shall refer to the conformation as $F(3_{1},3_{1})$. To
disentangle $F(3_{1},3_{1})$, one must slip the loops back all around the
bodies of the overhand knots, which is difficult, since the move needs first
making the loops bigger.

\bigskip \FRAME{ftbpFU}{4.625in}{3.4748in}{0pt}{\Qcb{Evolution of the lenght
of the rope in a process in which SONO disentagles the Freedman's $%
F(3_{1},3_{1})$ conformation of the unknot. Initially, the loose $%
F(3_{1},3_{1})$ conformation is rapidly tightened. Then, the evolution slows
down. At the end of the slow stage one of the end knots becomes untied.
Subsequently, the other of the end knots becomes untied. Eventually the
conformation becomes disentagled and the unknot reaches its ideal, circular
shape. The lower curve shows the evolution of the rope lenght in the much
faster process in which the unknot shown in Fig.1 becomes disentangled.}}{%
\Qlb{Fig2}}{fig2.jpg}{\special{language "Scientific Word";type
"GRAPHIC";maintain-aspect-ratio TRUE;display "USEDEF";valid_file "F";width
4.625in;height 3.4748in;depth 0pt;original-width 10.6666in;original-height
8.0004in;cropleft "0";croptop "1";cropright "1";cropbottom "0";filename
'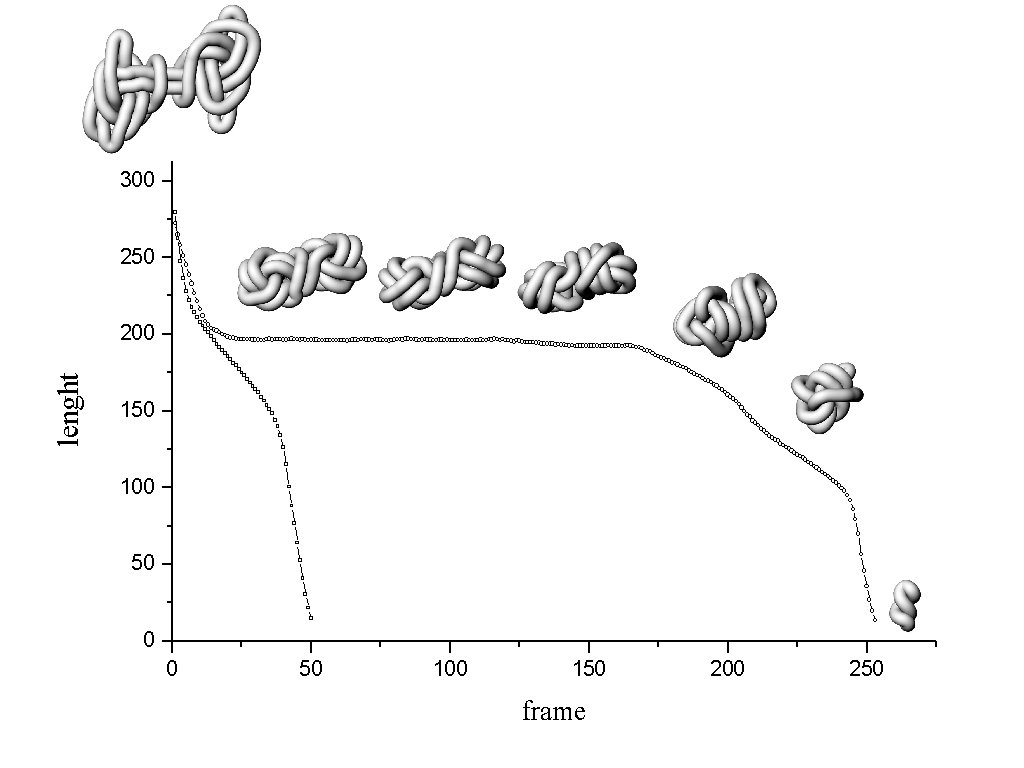';file-properties "XNPEU";}}

How the SONO algorithm copes with this task is shown in Fig.\ref{Fig2},
where consecutive stages of the disentangling process are shown. Tightening
the $F(3_{1},3_{1})$ conformation SONO algorithm brings it to the very
compact state, which seems at the first sight to be impossible to
disentangle. The end loops are very tight and they seem to be too small to
slip back over the bodies of the overhand knots. However, as the computer
simulations prove, there exists a path in the configurational space of the
knot along which the loops slowly become bigger and one of them slips over
the body of the overhand knot. Then, the disentangling process proceeds
without any problems. Results of the computer experiments we performed
suggest strongly, that the $F(3_{1},3_{1})$ conformation is not gordian.

The construction of original Freedman entanglement may be modified making it
more difficult to disentangle. The simplest way of doing this is to change
the end trefoil knots to some more complex knots. For the sake of brevity we
will use $F(K^{(1)},K^{(2)})$ symbols to indicate with what kind of the
Freedman conformation of the unknot we are dealing with. Results of computer
simulations we performed prove that the $F(4_{1},4_{1})$ conformation is
also disentangled in the knot tightening process. However, the $%
F(5_{1},5_{1})$ conformation proves to be resistant to SONO algorithm. Fig.%
\ref{Fig3} shows consecutive stages of the tightening process. The initial
conformation, is loose, it becomes tight soon. Then the evolution process
slows down and eventually stops. The final conformation is proves to be
stable. The gordian conformation has been reached.

\FRAME{ftbpFU}{3.9608in}{2.2451in}{0pt}{\Qcb{SONO tightens the $%
F(5_{1},5_{1})$ conformation of the unknot, but does not manage to
disentangle it.}}{\Qlb{Fig3}}{fig3.jpg}{\special{language "Scientific
Word";type "GRAPHIC";maintain-aspect-ratio TRUE;display "USEDEF";valid_file
"F";width 3.9608in;height 2.2451in;depth 0pt;original-width
9.8649in;original-height 5.5728in;cropleft "0";croptop "1";cropright
"1";cropbottom "0";filename '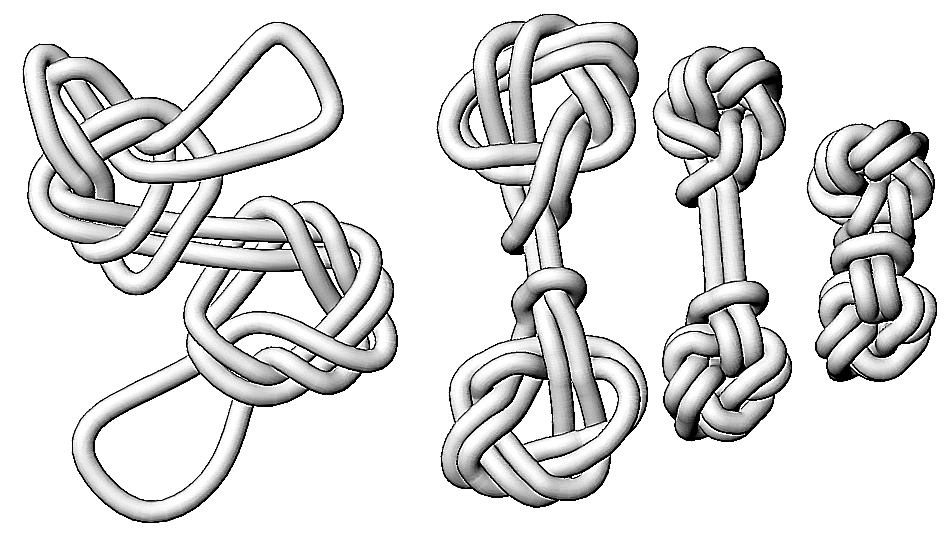';file-properties "XNPEU";}}

Eperimenting with knots tied on real, macroscopic ropes or tubes is by no
means easy \cite{Buck}. First of all, the surface of any real rope is never
smooth and strong friction often stops the walk within the configurational
space of a knot tied on such a rope. The role of friction was exposed by
Kauffman \cite{Kauffman}. Friction can be significantly reduced, however,
when a knot is tied on a smooth nanoscopic filament, e.g. a nanotube, or on
a thermally fluctuating polymer molecule \cite{DeGennes}. There exists
another, less obvious, factor which makes laboratory experiments on knots
difficult: the Berry's phase \cite{Berry}, to be more precise, its classical
counterpart - the Hannay's angle \cite{Hannay}. Modern ropes are often
constructed in the following manner: a parallel bundle of smooth filaments
is kept together by a tube-like, plaited cover. As easy to check, such ropes
are much easier to bend than to twist. Forming a knot on a rope, one has to
deform it. In view of what was said above, the deformation applied is rather
bending than twisting. Avoiding the twist deformations one follows the
procedure known as the parallel transport. As a result, when at the final
stage of the knot tying procedure the ends of the rope meet, they are in
general rotated in relation to each other: the misfit angle $A$ is the
Hannay's angle. As shown in \cite{Maggs} and \cite{Klapper}, the Hannay's
angle $A$ stays in a simple relation, 
\begin{equation*}
1+Wr=(A/2\pi )\func{mod}2
\end{equation*}
with the writhe $Wr$ of the knot into which the rope has been formed.
Splicing the ends of the rope one fixes the misfit angle $A$. Consequently,
the writhe value $Wr$ becomes fixed as well. As a result, any further
changes of the conformation of the knot become very difficult and are
basically restricted to the manifold of constant writhe. (The specific
construction of the Freedman conformations makes them achiral \cite{Liang}.
Their writhe is equal zero.)

The natural question arises, if the impossibility of disentangling the
gordian conformation does not stem from the described above friction and
writhe factors. We feel emphasize, that it is not the case. The rope
simulated by the SONO algorithm is perfect: it is frictionless and utterly
flexible. It has no internal, parallel bundle structure and it accepts any
twist. Problems with disentangling the gordian conformations are purely
steric. Tightening the $F(5_{1},5_{1})$ Freedman conformation SONO\ brings
it into a cul-de-sac of what mathematicians call thickness energy \cite
{Thick}. To get out of it, one needs elongate the rope. By how much? We do
not know yet the answer to this question.

We thank Jacques Dubochet, Giovanni Dietler, Kenneth Millett, Robert Kusner,
Alain Goriely, Eric Rawdon, Jonathan Simon, Gregory Buck and Joel Hass for
helpful discussions and correspondence. PP thanks the Herbette Foundation
for financial support during his visit in LAU. This work was carried out
under Project KBN 5 PO3B 01220.


\begin{thebibliography}{99}
\bibitem{Hass}  An equivalent prescription for creation of the Freedman
conformations of the unknot was formulated by Joel Hass (private
communication).
\end{thebibliography}
\end{document}